\documentclass{ws-ijbc}

\usepackage{graphicx}
\graphicspath{{pdf_figures/}}
\usepackage{multirow,bigstrut}
\usepackage{amssymb,amsmath,bm}

\newlength\figwidth
\usepackage{color}

\newlength\singleimagewidth
\newlength\twoimagewidth
\newlength\imagewidth
\begin{document}

\catchline{}{}{}{}{} 

\markboth{C. Li et al.}{Analysis of the state space of discrete dynamical systems}

\title{On the network analysis of the state space of discrete dynamical systems}

\author{Cheng Xu}
\address{School of Mathematics and Computational Science,\\
Xiangtan University, Xiangtan 411105, Hunan, China}

\author{Chengqing Li}

\address{College of Information Engineering,\\
 Xiangtan University, Xiangtan 411105, Hunan, China\\
 DrChengqingLi@gmail.com}

\author{Jinhu L\"u}
\address{Academy of Mathematics and Systems Sciences,\\
 Chinese Academy of Sciences, Beijing 100190, China}

\author{Shi Shu}
\address{School of Mathematics and Computational Science,\\
Xiangtan University, Xiangtan 411105, Hunan, China}

\maketitle

\begin{history}
Feb 15, 2017
\end{history}

\begin{abstract}
This paper discusses the letter entitled ``Network analysis of the state space of discrete dynamical systems" by A. Shreim et al. [Physical Review Letters, 98, 198701 (2007)]. We found that some theoretical analyses are wrong and the proposed indicators based on two parameters of the state-mapping network cannot discriminate the dynamical complexity of the discrete dynamical systems composed of a 1-D Cellular Automata.
\end{abstract}


\section{Introduction}

Methodology of complex networks has been used as an important tool to investigate the dynamics of some discrete systems and even chaotic cryptanalysis \cite{Liou:Network:IJMPB12,Kayama:networkCA:LNCS12,Sanchez:IJBC:2015,WangQX:HDDCS:TCAS2016,Cqli:Fridrich:SP2017}.
The general framework of the analyses can be summarized as four stages:
selection of nodes; establishment of linkage between any pair of nodes; definition of some characteristic parameters of the obtained network;
comparison with other recognized (convincing) metrics. Interestingly, some subtle dynamical properties of the underling
discrete systems were disclosed from the perspective of complex networks \cite{Andrew:CA:1992,Thurner:Logistic:ICCS2007,Luque:FeigenbaumChaos:PLOS11}.

In paper \cite{ShreimNetworkCA2007}, the authors analyzed the state-mapping network of discrete dynamical systems composed of a 1-D cellular
automata (CA) with rules involving two colors and the nearest neighbors. The studied CA sets the binary value $s^{t+1}_i$ of site $i$ in the $(t+1)$th iteration as $R(s^t_{i-1} s^t_i s^t_{i+1})$ counted at the previous one, where $R$ denotes the CA rule with an identifying number $R(000)2^0+R(001)2^1+R(010)2^2+R(011)2^3+R(100)2^4+ R(101)2^5+R(110)2^6+R(111)2^7$. Then, two characteristic parameters of the built directed network were adopted: the largest in-degree and
the \emph{path diversity} measuring fluctuations among different paths connecting the nodes corresponding to the transient states with zero in-degree
and that corresponding to attractors. Based on observation results on only ten rules, the authors claimed that ``the scaling and distribution of in-degrees and the path diversity give a good indication of dynamical complexity". More precisely, the statement means that the co-appearance of nontrivial scaling of both
in-degree distribution (or the hub sizes) and the path diversity of a state-mapping network with its system size can separate simple dynamics from the more complex ones found in CA falling in Wolfram's classes III and IV. This paper is to point out that the analytic results for rule~4 given in \cite{ShreimNetworkCA2007} are wrong and the proposed indicators cannot discriminate the dynamical complexity of CA at all.

\section{Discussion on network analysis of the state space of discrete dynamical systems}

\begin{figure}[!htb]
\centering
\includegraphics[width=0.9\figwidth]{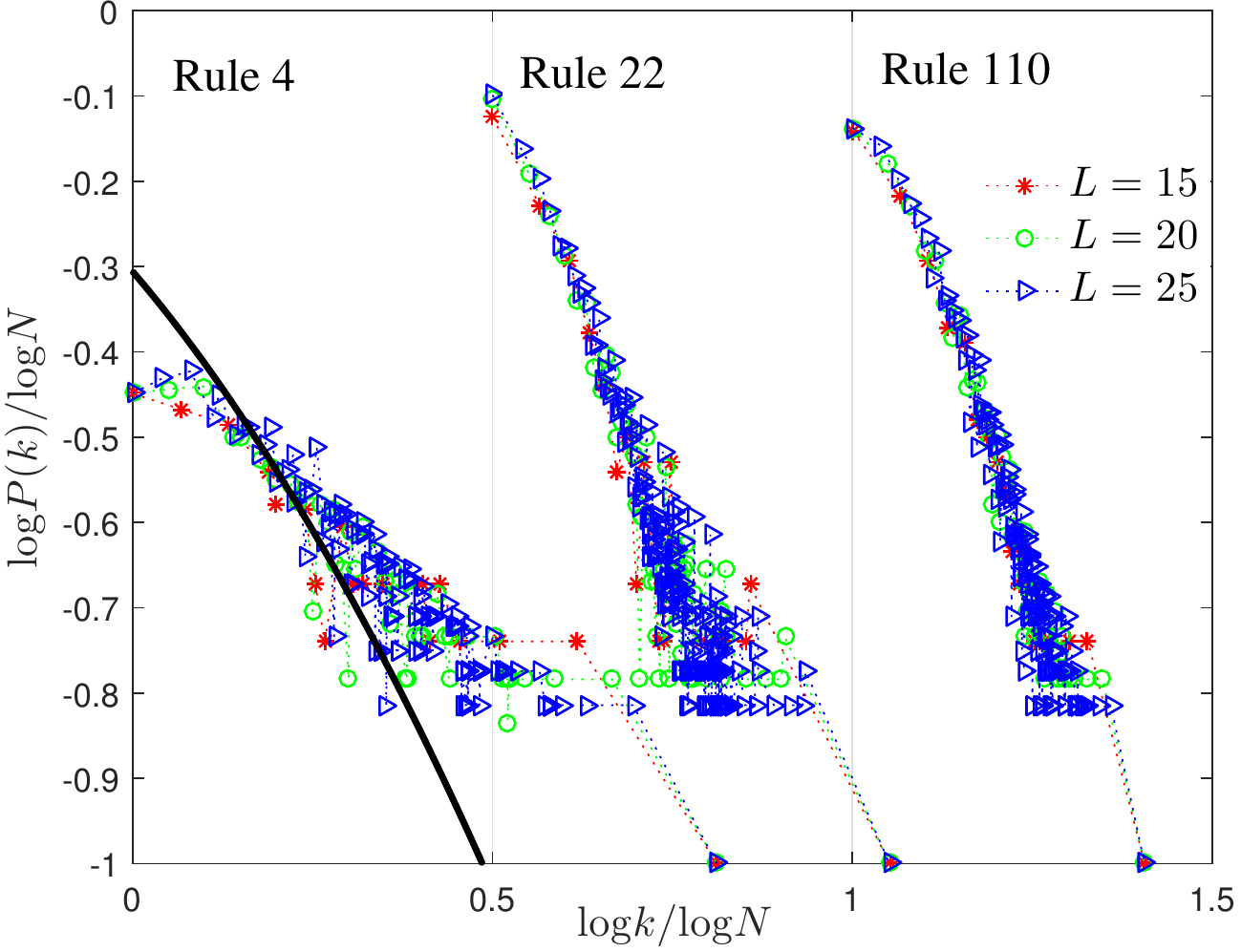}
\caption{In-degree distribution functions for three rules with different system sizes.}
\label{fig:InfluenceMatrix}
\end{figure}

In \cite{ShreimNetworkCA2007}, there are some obvious discrepancies on the estimation of in-degree distribution of the state-mapping network corresponding to rule~4.
The authors adopted $a\cdot\lambda^{m_i}$ as an approximate value of $$w(m_i)={\bf e}_3 \; [{\bf T}^{(0)}]^{m_i} \; {\bf e}_2^{\mathsf{T}},$$
where $\lambda$ is the largest eigenvalue of ${\bf T}^{(0)}$, a fixed binary matrix facilitating counting the in-degree of any state-mapping of the state-mapping network
of CA with rule~4,
${\bf e}_2 = (0,1,0,0)$, ${\bf e}_3 = (0,0,1,0)$, and $m_i$ is the number of `0's following the $i$th `1' in the counted state.
However, the fact is that the estimation errors are accumulated exponentially in the multiplications in $$k_n=\prod_{i=1}^n w(m_i)=\prod_{i=1}^n (a\lambda^{m_i}).$$
The authors of \cite{ShreimNetworkCA2007} calculate $\Omega(n)$, the number of states with $n$ isolated 1's but no pairs of `11', obtaining $C(L-n,n)+C(L-n-1,n-1)$. Actually, $\Omega(n)=C(L-n+1, n)=C(L-n,n)+C(L-n, n-1)$.
As for rule 4, every non-isolated node satisfies $L-n+1\ge n$. So, Eq.~(7) in \cite{ShreimNetworkCA2007},
\begin{equation}
   y \approx -1 - x + \log_2 \left[\frac{(1-\epsilon)^{1-\epsilon}}
         {\epsilon^\epsilon(1-2\epsilon)^{1-2\epsilon}}\right],
\label{eq:approximate}
\end{equation}
outputs complex numbers as $(1-2\epsilon)\le 0$ when $L \le 2n \le L+1$,
where $\epsilon\equiv n/L$. The cascading effect of these errors implies that Eq.~(\ref{eq:approximate}) cannot accurately approximate the distribution of the largest in-degree values. In addition, the scale of the Y-axis in Fig.~4 should be $[-1, 0]$, not $[-2, 0]$, which can be
deduced from $1/N\leq P(k) \leq1$ and $-1\leq \log P(k)/\log N \leq 0$. We re-calculated the in-degree distribution functions for three rules with three system sizes in Fig.~\ref{fig:InfluenceMatrix}
and re-drew Eq.~(\ref{eq:approximate}) at its left part by a bold black curve, which confirm our findings.

\begin{figure}[!htb]
\centering
\includegraphics[width=1.2\figwidth]{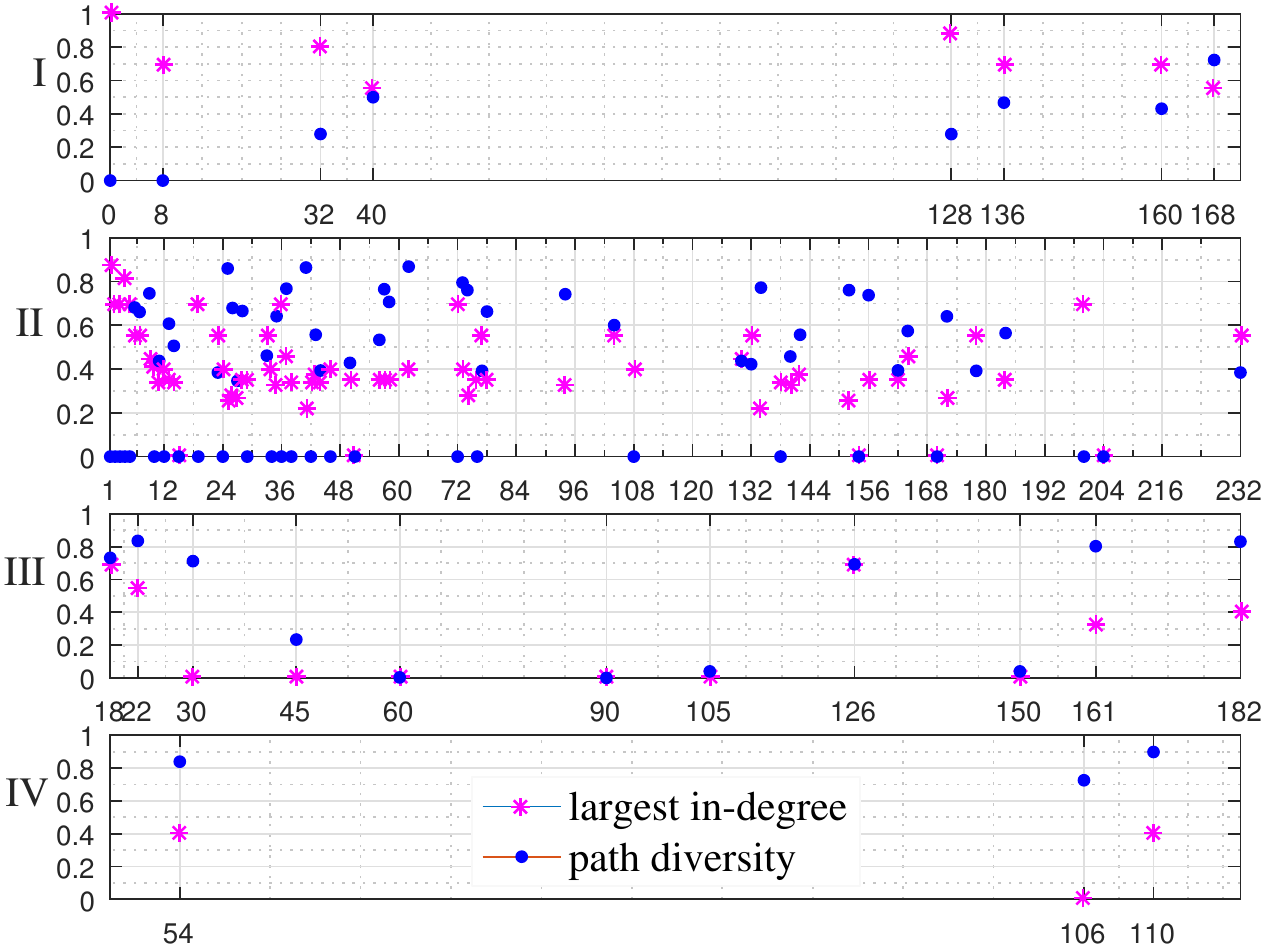}
\caption{Distribution of the two observables defined in \cite{ShreimNetworkCA2007} with respect to the rule numbers.}
\label{fig:classification}
\end{figure}

As to the 1-D CA under study, the 256 elementary rules are reduced to 88
independent ones under some transformations \cite[Page 294]{LWT:CA:CS90}. However, only ten independent rules were selected to verify the statement
in \cite{ShreimNetworkCA2007}. We calculated the change slope of the largest in-degree value and path diversity
with respect to the CA size for every independent rule and now depict the results in Fig.~\ref{fig:classification}, which is divided into
four panels according to Wolfram's classification (the detailed classification is referred to \cite[Page 231]{Wolfram:CA:2002}, \cite[Table 1]{LWT:CA:CS90}, and \cite[Table~3]{zenil2013asymptotic}).
From each panel, no general pattern can be observed. Furthermore, one can see that both the two observed values are very close to each other among rules 54, 62, 110, and 182, which belong to three different classes.
More counterexamples can be given to disprove inefficiency of the main statement of \cite{ShreimNetworkCA2007} even just based on Fig.~\ref{fig:classification}.

\section{Conclusion}

This paper briefly reported three key errors in the deduction process of estimating an in-degree distribution function in \cite{ShreimNetworkCA2007}.
With the help of Wolfram's classification on CA, we further demonstrated that the in-degree distribution and the path diversity of a state-mapping network cannot be used to measure the dynamical complexity of the corresponding underlying system formed by a 1-D CA. Justification for measuring complexities of general discrete dynamical systems using statistical characteristics of their state-mapping networks needs further investigation.
More detailed analysis on the state-mapping networks of 1-D CAs will be reported in a forthcoming paper.

\section*{Acknowledgement}

This research was supported by the Natural Science Foundation of China (No.~61532020, 61611530549) and Scientific
Research Fund of Hunan Provincial Education Department (No.~15A186).

\bibliographystyle{ws-ijbc}
\bibliography{CA_Letter_IJBC}
\end{document}